\documentstyle[emulateapj,psfig]{article}
\def\gs{\mathrel{\raise0.35ex\hbox{$\scriptstyle >$}\kern-0.6em 
\lower0.40ex\hbox{{$\scriptstyle \sim$}}}}
\def\ls{\mathrel{\raise0.35ex\hbox{$\scriptstyle <$}\kern-0.6em 
\lower0.40ex\hbox{{$\scriptstyle \sim$}}}}
\slugcomment{Submitted to The Astrophysical Journal (Letters)}

\lefthead{Brainerd \& Specian}
\righthead{Mass--to--Light Ratios of 2dF Galaxies}

\begin{document}

\title{Mass--to--Light Ratios of 2dF Galaxies}

\author{Tereasa G.\ Brainerd \& Michael A.\ Specian}
\affil{\tiny Boston University, Department of Astronomy, 725 Commonwealth Ave., 
Boston, MA 02215}

\begin{abstract}
We compute
$M_{260}^{\rm dyn}$,  the dynamical mass interior to a radius of 
$260 h^{-1}$~kpc, for a set of 809 isolated host
galaxies in  the 100k data release of the 2dF Galaxy Redshift
Survey.  The hosts are surrounded by 1556 satellite
galaxies, as defined by a set of specific selection criteria.  
Our mass estimator and host/satellite selection criteria are
taken from those used by the Sloan Digital Sky Survey (SDSS)
collaboration for an analysis of
$M_{260}^{\rm dyn}$ for SDSS galaxies and, overall,
our results compare well with theirs.  In particular, for
$L \gs 2L^\ast$ we find $(M_{260}^{\rm dyn}/L)_{b_J} =  
(193 \pm 14) h~M_\odot/L_\odot$, 
with a weak tendency
for hosts with $L < 2L^\ast$ to have a somewhat higher $M/L$.  Additionally, we
investigate $M/L$ for bright ($b_J \ls 18$) galaxies with elliptical, S0, and
spiral morphologies.  There are 159 hosts in the elliptical/S0 sample and, similar
to the full sample, we find $(M_{260}^{\rm dyn}/L)_{b_J} = 
(271 \pm 26) h~M_\odot/L_\odot$ 
for galaxies
with $L \gs 2L^\ast$, and a weak tendency for intrinsically fainter galaxies to
have a somewhat higher $M/L$.  In stark contrast to this, we find the line of
sight velocity dispersion for the 243 spiral hosts to be independent of the
host luminosity, with a value of $\sigma_v = 189\pm 19$~km~s$^{-1}$.  
Thus, for
spiral hosts we find that 
$(M_{260}^{\rm dyn}/L)_{b_J} \propto L^{-1.0\pm 0.2}$, 
where $(M_{260}^{\rm dyn}/L)_{b_J}$ for a $2L^\ast$ spiral 
is of order $200 h~M_\odot/L_\odot$.

\end{abstract}



\section{Introduction}

It is generally accepted that large, bright galaxies reside within
massive dark matter halos; however, the radial extent of the halos
is not well--constrained
and, hence, neither is the total mass nor the mass--to--light
ratio of these objects.  Galaxy--galaxy lensing, in which the
halos of foreground galaxies weakly distort the shapes of background
galaxies, has recently proven to be a powerful method by which the masses and
mass--to--light ratios of galaxies may be constrained.  Galaxy--galaxy
lensing has been detected by a number of different groups (see, e.g., 
the review by
Brainerd \& Blandford 2003 and references therein) and, in particular, the
Sloan Digital Sky Survey (SDSS) collaboration has obtained measurements
of the galaxy--galaxy lensing shear with extremely high statistical
significance (e.g., Fischer et al.\ 2000; McKay et al.\ 2001).

Using weak lensing
measurements of the projected mass correlation function, McKay et al.\
(2001), hereafter SDSS01, found that $M_{260}^{\rm lens}$,
 the mass of lens galaxies interior to a radius of
$260 h^{-1}$~kpc, scaled roughly linearly with the luminosities
of the lens galaxies in all bandpasses except $u'$.   Since the 
galaxy--galaxy lensing shear is small ($\ls 0.5$\% in the case of the SDSS
galaxies) and is not without its own sources of error (including the 
the separation of lenses from sources), McKay et al.\ (2002) performed
an independent estimate of the masses of dark matter halos surrounding SDSS
galaxies using the dynamics of satellite galaxies.  Their sample consisted
of 618 host galaxies and 1225 satellites,  which was considerably
smaller and shallower
than the sample in the weak lensing analysis due to the necessity of
redshifts for all of the galaxies.  Nevertheless, McKay et al. (2002), hereafter
SDSS02, found that their dynamical analysis led to trends in the dependence
of $M_{260}^{\rm dyn}$ on the host galaxy luminosity  that were reasonably
consistent with the trends obtained from their previous weak lensing 
analysis.  However, the mass--to--light ratios found from the dynamical
analysis were systematically lower than those from the
lensing analysis ($M_{260}^{\rm dyn}/L \sim 0.8 M_{260}^{\rm lens}/L$).

Here we perform a dynamical analysis of the masses of isolated host galaxies
in the 100k public data release of the 2dF Galaxy Redshift Survey, hereafter
2dFGRS.
The 2dFGRS is a spectroscopic survey in which the objects are selected in
the $b_J$ band from the APM galaxy survey (Maddox et al.\ 1990a,b), and 
extensions to the original survey.  Ultimately, the survey will provide
spectra for $\sim 250,000$ galaxies brighter than $b_J = 19.45$ and will
cover an area of order 2000 square degrees (see, e.g., Colless et al.\ 2001).

Our host galaxies span
a redshift range which is similar to that of the SDSS02 galaxies, and our
sample is of a similar size.  We select host/satellite combinations and 
determine dynamical masses for the host galaxies based upon the methods outlined in
SDSS02 in order to compare most easily to their results.  In particular, we
investigate the apparent lack of dependence of $M_{260}^{\rm dyn}/L$ on the
host luminosity found by SDSS02, and the somewhat low value of the
dynamical mass--to--light ratio 
in comparison to the lensing mass--to--light ratio.

Throughout, we adopt a flat, $\Lambda$--dominated universe with
parameters $\Omega_0 = 0.3$, $\Lambda_0 = 0.7$, and 
$H_0 = 100h$~km~s$^{-1}$~Mpc~$^{-1}$.  Consistent with this, we take the
absolute magnitude of an $L^\ast$ galaxy in the $b_J$ band to be
$M_{b_J}^\ast- 5 \log_{10} h = -19.66 \pm 0.07$ (Norberg et al., 2002).

\section{Host and Satellite Selection}

In order to compare to the results of SDSS02, we 
select host and satellite galaxies from the 2dF survey according to the
SDSS02 criteria:

\begin{itemize}
\item Host galaxies must be ``isolated''.  They must be
at least twice as luminous as any other galaxy that falls within a projected radius of
$2h^{-1}$~Mpc, as well as within a velocity difference of
$|dv| \le 1000$~km~s$^{-1}$.
\item Potential satellite galaxies must be at least 4 times fainter
than their host, must fall within a projected radius of $500 h^{-1}$~kpc of their
host, and
the satellite--host velocity difference must be $|dv| \le 1000$~km~s$^{-1}$.
\end{itemize}

These basic selection criteria result in 864 hosts and 2340 satellites.  As
noted by SDSS02, however, many of the hosts have a large number of satellites
around them (in one case, a potential host in our sample has 605 satellites).
These
are, therefore, most likely to be associated with cluster systems, rather than
being truly isolated.  To eliminate these objects, we impose a further restriction
that the luminosity of the host be greater
than the sum total of the luminosities of
the satellites.  This, too, was done by SDSS02, and reduces our 2dF sample to
859 hosts and 1693 satellites.

%
%
\vspace*{-0.5truecm}
\hbox{~}
\centerline{\psfig{file=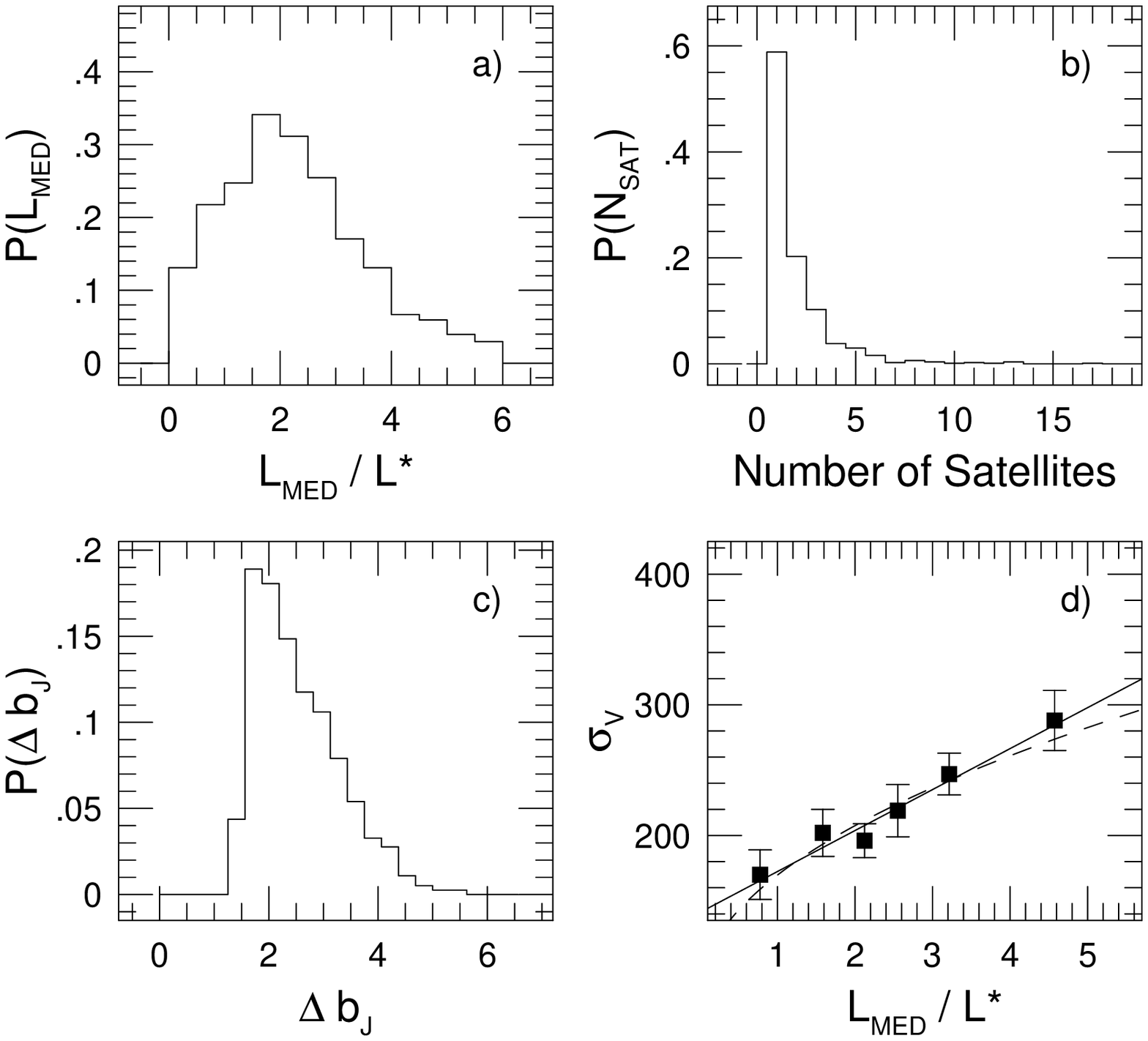,angle=0,width=4.0in}}
\vspace*{-4.5truecm}
\noindent{\scriptsize
\addtolength{\baselineskip}{-3pt} 
Figure~1. a) Probability distribution for luminosities of the 809 host galaxies
in the full sample.  Luminosities are in units of $L^\ast$. b) Probability 
distribution for the number of satellites surrounding the host galaxies in
the full sample.
c) Probability distribution for the difference in apparent magnitude between
the host and satellite galaxies in the full sample.
d) Velocity dispersion of satellite galaxies in the full sample as a function
of median host luminosity.  Solid line shows $\sigma_v \propto L$, which
is the best fit to the data.  Dashed line shows
$\sigma_v \propto \sqrt{L}$.
\label{fig1}
}

\bigskip
Finally, we impose two additional cuts on the host galaxies.  First, eyeball
morphologies are available for the 2dF galaxies with $b_J \ls 18$, and 29
of the above hosts are classified as galaxy--galaxy mergers.  We delete these hosts
from the sample on the basis that they are unlikely
to be fully relaxed systems.  Second, we delete all hosts with $L > 6 L^\ast$
because the velocity dispersions of their satellites are poorly fit
by the technique we adopt (see below), and the number of interloper galaxies
(as opposed to genuine satellites) appears to be both large ($\gs 45$\%) and
have a large dispersion ($\sim 20$\%).  These additional cuts leave us with 
a final
sample of 809 host galaxies and 1556 satellites.  Of these, 75 are classified
as ellipticals, 84 are classified as S0, and 243 are classified as spirals.
The sample of spirals is uniformly distributed in inclination angle, and there
is no correlation between host luminosity and median inclination angle.
The ellipticals have a total of 171 satellites, the S0's have a total of 303 
satellites, and the spirals have a total of 478 satellites.  The median 
redshift of the 809 host galaxies in the full sample is $z_{\rm med}= 0.073$, while
for the spiral hosts $z_{\rm med} =  0.055$, and for the elliptical and
S0 hosts $z_{\rm med} = 0.062$.

%
%
\vspace*{-0.5truecm}
\hbox{~}
\centerline{\psfig{file=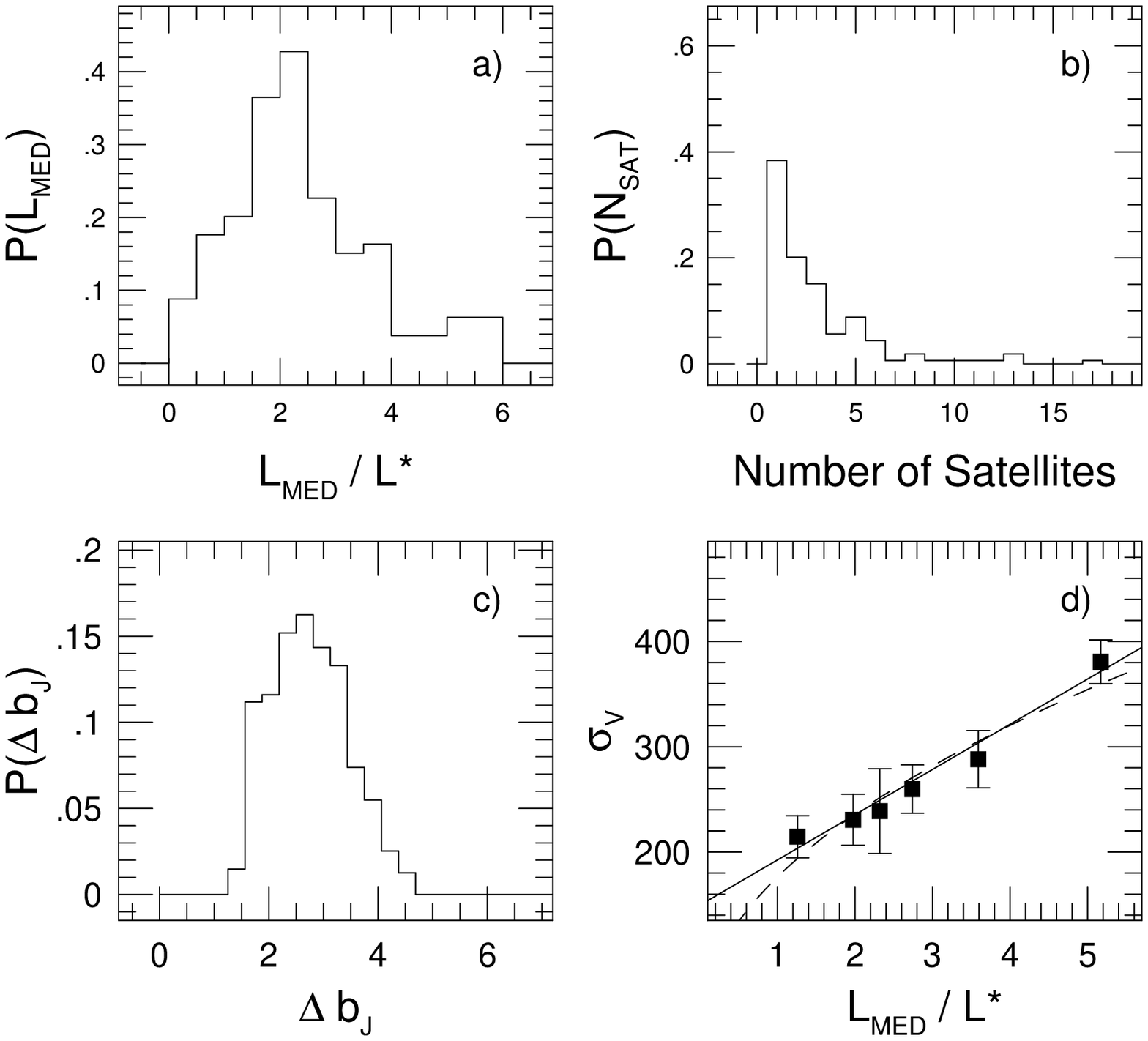,angle=0,width=4.0in}}
\vspace*{-4.5truecm}
\noindent{\scriptsize
\addtolength{\baselineskip}{-3pt} 
Figure~2. Same as Fig.\ 1, but for elliptical and S0 host galaxies.
\label{fig2}
}

\bigskip
The probability distribution of the luminosities of the host galaxies, the 
probability distribution of the number of satellites around individual hosts,
and the probability distribution of the difference in apparent $b_J$ magnitude
between the hosts and their satellites are shown in panels a, b, and c of
Figs.\ 1, 2, and 3.  Fig.\ 1 shows results for the entire sample of 809 hosts,
while Fig.\ 2 shows the results for the 159 hosts classified as elliptical 
or S0, and Fig.\ 3 shows the results for the 243 hosts classified as spirals.

\section{Halo Velocity Dispersions}

In order to compare with SDSS02, we adopt an analysis technique
that is identical to theirs.  The radial velocity dispersions of the
host galaxy halos, $\sigma_v$, are computed by fitting a combination of a
Gaussian and a constant offset to histograms of the velocity differences
between the hosts and satellites.  The width of the best--fitting Gaussian
is a measure of $\sigma_v$, while the offset accounts for the fact that there
will, necessarily, be some fraction of interloper galaxies that are selected
as satellites when, in fact, they are not dynamically associated with the
host galaxy.  Like SDSS02, 
we find that this technique provides
very good fits to the velocity difference histograms, yielding values of
$\chi^2$ per degree of freedom, $\chi^2/\nu$, that are $\ls 1$ for hosts with 
$L \le 6L^\ast$.  In the case of hosts with $L > 6L^\ast$, $\chi^2/\nu \gs 2.5$
and, hence, we do not consider these objects further. 

Because it is likely that more interlopers will have velocities that are
greater than their hosts (e.g., Zartisky \& White 1994), 
we determined $\sigma_v$
for the host galaxies by fitting Gaussians plus constant offsets to 3 different
velocity difference histograms: (i) velocity differences taken to be
the absolute value, $|dv|$, of the measured difference, (ii) negative velocity
differences, and (iii) positive velocity differences.   We define the velocity
difference to be $dv \equiv v_{\rm host} - v_{\rm sat}$, so that negative
values of $dv$ correspond to satellites which are more distant in velocity
space.  In all cases, the best--fitting velocity dispersions are in very good
agreement amongst the 3 histograms.  In addition, we find a clear difference
in the number of interlopers.  In the case of the full sample of 809 hosts,
the interloper fraction is $(31\pm 3)$\% for host--satellite pairs with
$dv < 0$, $(20 \pm 3)$\% for host--satellite pairs with $dv > 0$, and 
$(27\pm 2)$\% on average (i.e., fitting to the distribution of $|dv|$).  
For the hosts with spiral morphologies, the mean interloper fraction is
$(33\pm 3)$\% while for hosts with elliptical and S0 morphologies the
interloper fraction is much lower, $(14\pm 4)$\%.
Lastly, for all of our samples of host galaxies,
we find that $\sigma_v$ is independent of the radius at which
it is determined.

%
%
\vspace*{-0.5truecm}
\hbox{~}
\centerline{\psfig{file=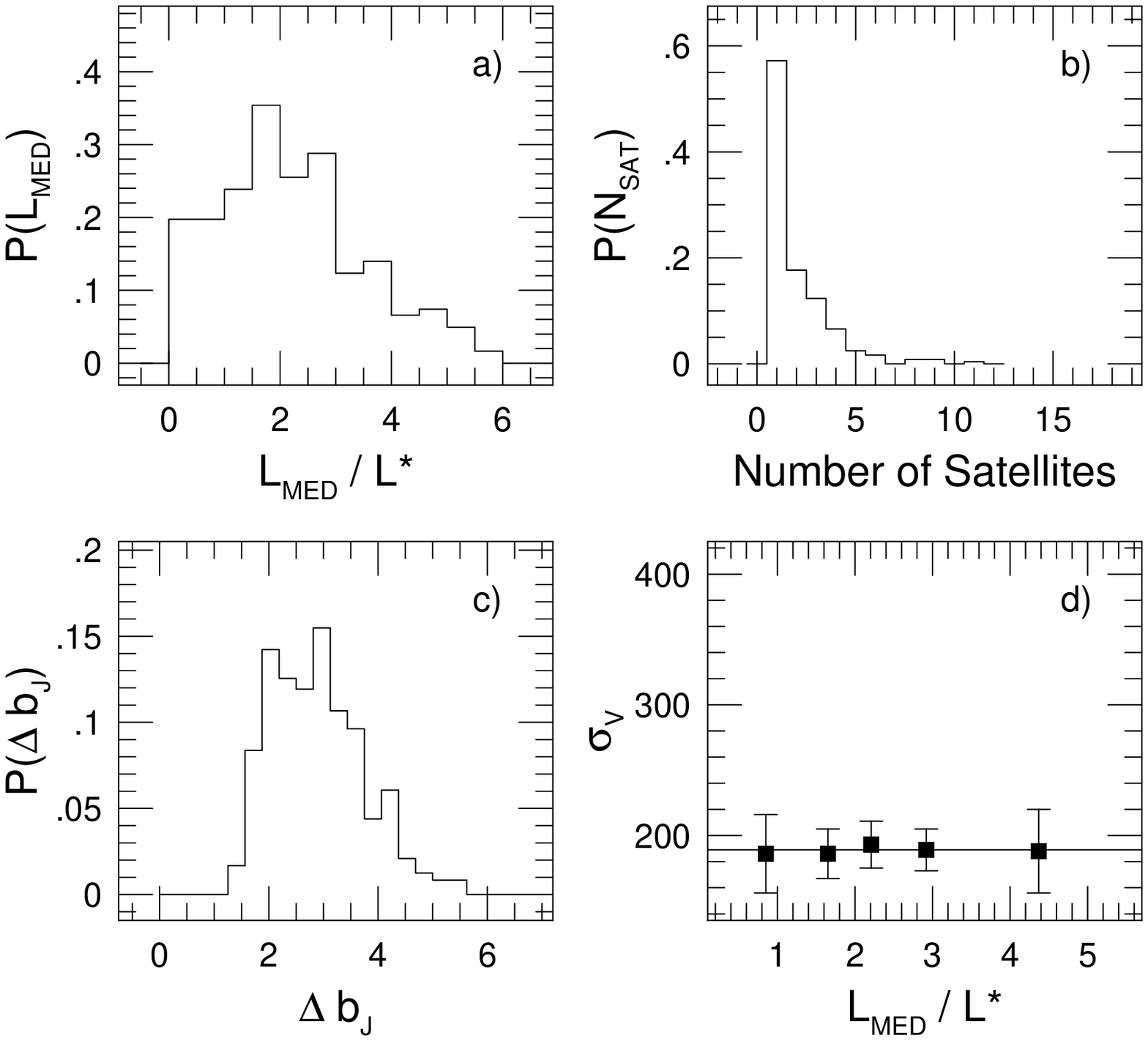,angle=0,width=4.0in}}
\vspace*{-4.5truecm}
\noindent{\scriptsize
\addtolength{\baselineskip}{-3pt} 
Figure~3. Same as Fig.\ 1, but for spiral host galaxies.  Solid line in
panel d shows $\sigma_v = 198$~km~sec$^{-1}$, which is the best fit to
the data.
\label{fig3}
}

\bigskip
Results for $\sigma_v$ as a function of median host luminosity 
are shown in panel d of Figs.\ 1, 2, and 3.  In the case of the full host
sample and the elliptical/S0 host sample, the relationship between velocity
dispersion and median host luminosity  is fit best by linear relations,
and these are shown by the solid lines in panel d of Figs.\ 1 and 2.  
For the full host sample we find 
$\sigma_v = (31\pm 8)L/L^\ast + (141\pm 18)$~km~s$^{-1}$,
and for the elliptical/S0 sample we find $\sigma_v = (43\pm 7)L/L^\ast
 + (149\pm 22)$~km~s$^{-1}$. This
is somewhat different from the results of SDSS02 who found $\sigma_v \propto
\sqrt{L}$.  We note, however, that our data are consistent with 
such a relationship,
and the best--fitting function of the form $\sigma_v \propto \sqrt{L}$ is
shown by the dashed lines in these figures.  Strikingly different from these
results, however, is the relationship of velocity dispersion and median host
luminosity for the spiral hosts.  From panel d of Fig.\ 3, it is clear that
$\sigma_v$ for the spiral hosts is independent of host luminosity, and we find
$\sigma_v = 189\pm 9$~km~s$^{-1}$ for these objects.

\section{Host Masses and Mass--to--Light Ratios}

To determine the masses of the dark matter halos which surround our host
galaxies, we adopt the following mass estimator:

\begin{equation}
M(r) = - \frac{ r \left< v^2_r \right> }{G} 
\left[
\frac{\partial \ln \rho}{\partial \ln r} +
\frac{\partial \ln \left< v^2_r \right>}{\partial \ln r} +
2\beta
\right]
\end{equation}

\noindent
(e.g., Binney \& Tremaine, 1987).  Here $r$ is a 3-dimensional radius,
$\left< v^2_r \right>$ is the mean
square radial velocity of the satellites, $\rho(r)$ is the number density of
satellites, and $\beta$ is a measure of the anisotropy in the velocity dispersion
of the satellites:

\begin{equation}
\beta \equiv 1 - \frac{ \left< v^2_\theta \right> }{ \left< v^2_r \right> } .
\end{equation} 

Although other methods of obtaining dynamical masses using
satellite galaxies have been adopted in
the literature (see, e.g., Bahcall \& Tremaine 1982; Zaritsky \& White 1994;
Zaritsky et al.\ 1997),
this is the method adopted by SDSS02 and, therefore, we adopt it as well.  
SDSS02
have used the GIF simulation, which incorporates semi--analytic
galaxy formation within a large cosmological N--body simulation (e.g., 
Kauffmann et al.\ 1999), 
to evaluate this mass estimator.  In particular, SDSS02
find that the velocity anisotropy of the satellite galaxies in the GIF 
simulation is small (i.e., $\beta$ is consistent with zero at the 2--$\sigma$
level), and that the mean square line of sight velocity dispersion, 
$\sigma_v^2$,
is consistent with the mean square radial velocity dispersion, 
$\left< v^2_r \right>$, at the 1--$\sigma$ level.  Combining this with
the fact that the line of sight velocity dispersion is observed to be
independent of radius, the mass estimator
used by SDSS02 reduces to:

\begin{equation}
M(r) = - \frac{ r \sigma_v^2 }{G}
\frac{\partial \ln \rho}{\partial \ln r} .
\end{equation}

We compute $\rho(r)$ for the satellites in our sample and find
$\rho(r) \propto r^{-2.11\pm 0.06}$ for the satellites surrounding the
hosts in the full sample, $\rho(r) \propto r^{-2.11\pm 0.09}$ for
the satellites surrounding elliptical and S0 hosts, and 
$\rho(r) \propto r^{-2.2 \pm 0.1}$ for the satellites surrounding 
spiral hosts.  These are all consistent with the results of SDSS02
who find $\rho(r) \propto r^{-2.1}$ for their sample.

Having obtained the number density of satellites as a function of
radius, we now use the values of $\sigma_v$ from Figs.\ 1, 2, and
3 to determine the mass--to--light ratios of the host galaxies.  
For consistency with SDSS02, we adopt a fiducial projected radius
of $260h^{-1}$~kpc.  Shown in Fig.\ 4 are the results for
$(M_{260}^{\rm dyn} /L)_{b_J}$ in units of $h~M_\odot/L_\odot$
as a function of median host luminosity.  The top panel shows results
for the full sample of host galaxies, the middle 
panel shows results for the elliptical and S0 hosts, and the bottom
panel shows results for the spiral hosts.

From Fig.\ 4, then, we find that $(M_{260}^{\rm dyn}/L)_{b_J}$ for the 809
hosts in our full sample is fairly constant for hosts with $L \gs 2L^\ast$ and
has a value of $(193\pm 14) h~M_\odot/L_\odot$.  For hosts with $L < 2L\ast$ there
is a weak suggestion of a somewhat higher mass--to--light ratio.  Similarly,
$(M_{260}^{\rm dyn}/L)_{b_J}$ for the elliptical and S0 hosts is fairly constant
for hosts with $L \gs 2L^\ast$ and has a value of $(271 \pm 26) h~M_\odot/L_\odot$.
Again, there is a slight suggestion that elliptical and S0 hosts with 
$L < 2L^\ast$ have a somewhat higher mass--to--light ratio.
In contrast, over the range of host luminosities explored here,
$(M_{260}^{\rm dyn}/L)_{b_J}$ for the spiral hosts
shows a clear monotonic decrease with luminosity, and is consistent with
a power--law of the form $(M_{260}^{\rm dyn}/L)_{b_J} \propto L^{-1.0 \pm 0.2}$.

%
%
\vspace*{-0.5truecm}
\hbox{~}
\centerline{\psfig{file=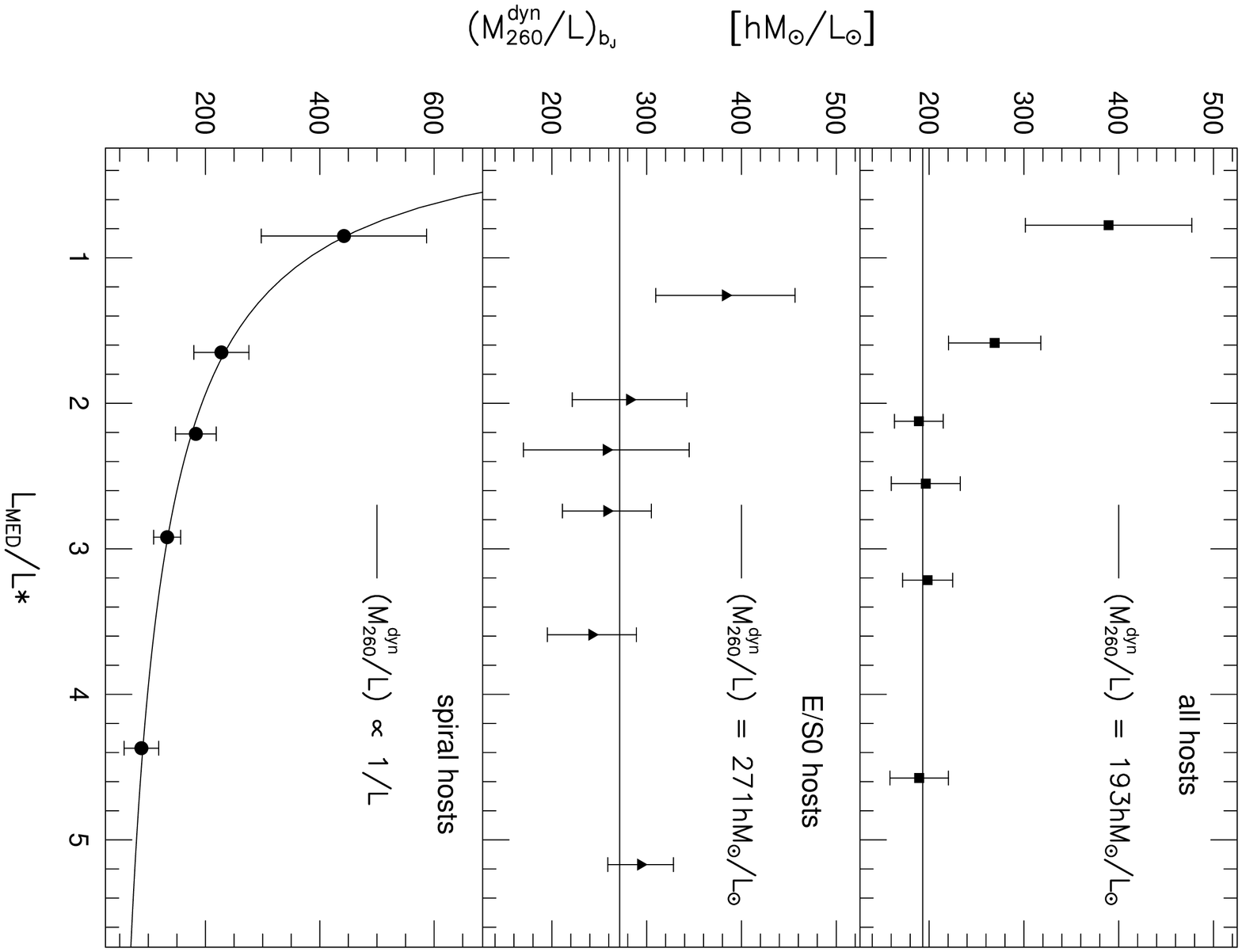,angle=-90,width=3.5in}}
\vspace*{1.5truecm}
\noindent{\scriptsize
\addtolength{\baselineskip}{-3pt} 
Figure~4. Mass--to--light ratios for the host galaxies interior to a radius
of $260 h^{-1}$~kpc.  Top panel: all 809 host
galaxies, middle panel: hosts with elliptical and S0 morphologies, bottom
panel: hosts with spiral morphologies.  
Solid lines in the top and middle panels show the mean mass--to--light
ratio for hosts with $L \gs 2L^\ast$.  Solid line in the bottom panel
show the best--fitting power law for $(M_{260}^{\rm dyn}/L)_{b_J}$.
\label{fig4}
\addtolength{\baselineskip}{3pt}
}

\section{Discussion}

The 2dF and SDSS02 host/satellite samples are of comparable depths and have
similar sizes, so it is not unreasonable to make comparisons between them.
The comparison 
is, however, somewhat limited
by the fact that the 2dF galaxies are selected in $b_J$, while the
SDSS02 galaxies are selected in $r'$, with host luminosities obtained
in $u'$, $g'$, $r'$, $i'$, and $z'$.  (The transformation from the SDSS
photometry is given by $b_J = g' + 0.155 + 0.152(g'-r')$; Norberg et al.\
2002).  Also,
SDSS02 did not perform separate dynamical analyses for the hosts of early-- and
late--type galaxies, so a direct comparison is not possible in this case.

In all 5 SDSS photometric bands, SDSS02 find $M_{260}^{\rm dyn} \propto L$, so
that in a given band, a single mass--to--light ratio characterizes the hosts.
That mass--to--light ratio is a sharply decreasing function of the central 
wavelength of
the bandpass (e.g., a factor of order 3 higher in $u'$ 
than in $z'$).  In $g'$, SDSS02 find 
$M_{260}^{\rm dyn}/L = (171 \pm 40) h~M_\odot/L_\odot$ and in $r'$
$M_{260}^{\rm dyn}/L = (145 \pm 34) h~M_\odot/L_\odot$.  These compare
well with the mass--to--light ratio that we obtain, $(193\pm 14) 
h~M_\odot/L_\odot$, 
for the host
galaxies our full sample that have luminosities of $L \gs 2L^\ast$.

Since we cannot compare our $M_{260}^{\rm dyn}/L$ for host galaxies of
different morphologies to the results of SDSS02, we instead compare them
to the weak lensing results of SDSS01.  SDSS01 did not classify their
galaxies according to visual morphology but, instead, used spectral features
to place subsets of their lens galaxies into broad ``early--'' and 
``late--type'' categories.  The early--types represent about 40\% of the
total number of lens galaxies, and the late--types represent another
40\% of the total number of lens galaxies.  Table 3 of SDSS01 shows that
in the bluer bands, $M_{\rm 260}^{\rm lens}/L$ 
is somewhat morphology--dependent,
with the mass--to--light ratio of the ellipticals exceeding that of
the entire lens sample by a factor of $1.5\pm 0.2$ in $g'$ and by a factor
of $1.3\pm 0.2$ in $r'$.  Again, this compares well with our results
for the elliptical/S0 hosts in the 2dF sample, where we find that
$M_{260}^{\rm dyn}/L$ for the elliptical/S0 hosts exceeds that of the full
sample by a factor of $1.4\pm 0.2$ for hosts with $L \gs 2L^\ast$
(e.g., Fig.\ 4).  

Our result that $M_{260}^{\rm dyn}/L \propto L^{-1}$ for the spiral hosts
is in clear conflict with the results of SDSS01, who found that
$M_{260}^{\rm lens}/L$ was independent of luminosity in all but the very
bluest band ($u'$).   However, our result stems from the fact that the
line of sight velocity dispersion is independent of luminosity for the
spiral hosts.  While this is inconsistent with the lensing results of SDSS01,
it is consistent with the dynamical results of Zaritsky et 
al.\ (1997) who found that the velocity difference, $dv$, between
69 isolated spiral galaxies ($-22.4 < M_B < -18.8$) and 115 satellites
was independent of the inclination--corrected
H-I linewidth of the host and was, therefore, 
independent of the
luminosity of the host (through, e.g., the Tully--Fisher relation).

Whether the conflict between the lensing and dynamical results for
the halos of spiral galaxies is due to differences in sample selection
or due to systematic effects in one or both of the mass estimators remains
to be determined.  However, the ultimate
completion of both the SDSS and the 2dFGRS will aid tremendously in
the resolution of this issue, and we look forward to the wealth of
data that both surveys will provide in the near future.

\section*{Acknowledgments}

We are pleased to 
thank the 2dFGRS team for making the 100k public data 
release available in a timely and very user--friendly manner.
Support under NSF contract AST-0098572 (TGB, MJS)
is also gratefully acknowledged.


\begin{references}

\reference{} Bahcall, J. N. \& Tremaine, S. 1981, ApJ, 244, 805

\reference{} Binney, J. \& Tremaine, S. 1987, Galactic Dynamics (Princeton:
Princeton Univ.\ Press)

\reference{} Brainerd, T. G. \& Blandford, R. D. 2003 in {\it Gravitational
Lensing: An Astrophysical Tool}, Springer Lecture Notes in Physics
vol.\ 608, eds.\ F.\ Courbin \& D.\ Minniti, 96

\reference{} Colless, M. M. et al.\ 2001, MNRAS, 328, 1039

\reference{} Fischer, P. et al.\ 2000, AJ, 120, 1198

\reference{} Maddox, S. J., Efstathiou, G., Sutherland, W. J. \& Loveday, J. 1990a,
MNRAS, 243, 692

\reference{} Maddox, S. J., Efstathiou, G., Sutherland, W. J. \& Loveday, J. 1990b,
MNRAS, 246, 433

\reference{} McKay, T. A. et al.\ 2001, ApJ submitted, astro--ph/0108013
(SDSS01)

\reference{} McKay, T. A. et al.\ 2002, ApJ, 571, L85 (SDSS02)

\reference{} Norberg, P. et al.\ 2002, MNRAS, 336, 907

\reference{} Kauffmann, G., Colberg, J. M., Diaferio, A. \& White,
S. D. M. 1999, MNRAS, 303, 188

\reference{} Zaritsky, D. \& White, S. D. M. 1994, ApJ, 435, 599

\reference{} Zaritsky, D., Smith, R., Frenk, C. \& White, S. D. M.,
1997, ApJ, 478, 39

\end{references}
\end{document}